\documentclass[runningheads]{llncs}
\usepackage{graphicx}
\usepackage{amsmath}
\usepackage{amssymb, placeins}
\usepackage{float}
\usepackage{lscape}
\usepackage{graphicx}
\usepackage[caption=false]{subfig}
\usepackage{booktabs}
\usepackage[ruled, linesnumbered]{algorithm2e}
\usepackage{setspace}
\usepackage{stackengine}
\usepackage{flushend}
\usepackage{multirow}
\usepackage{tikz}
\usepackage{graphbox}
\usepackage{gensymb}

\usetikzlibrary{arrows}
\tikzstyle{block} = [rectangle, draw, fill=blue!20, outer sep=5pt,
text width=4em, text centered, rounded corners, minimum height=4em]
\begin{document}
\title{Non-Iterative Phase Retrieval With Cascaded Neural Networks}
%
%
\author{Tobias Uelwer\and
Tobias Hoffmann \and
Stefan Harmeling}
\authorrunning{T. Uelwer et al.}
%
\institute{Department of Computer Science, Heinrich Heine University Düsseldorf, Germany\\
	\email{\{tobias.uelwer, tobias.hoffmann, stefan.harmeling\}@hhu.de}}
\maketitle              
\begin{abstract}
Fourier phase retrieval is the problem of reconstructing a signal given only the magnitude of its Fourier transformation.
Optimization-based approaches, like the well-established Gerchberg-Saxton or the hybrid input output algorithm, struggle at reconstructing images from magnitudes that are not oversampled.
This motivates the application of learned methods, which allow reconstruction from non-oversampled magnitude measurements after a learning phase. 
In this paper, we want to push the limits of these learned methods by means of a deep neural network cascade that reconstructs the image successively on different resolutions from its non-oversampled Fourier magnitude. 
We evaluate our method on four different datasets (MNIST, EMNIST,
Fashion-MNIST, and KMNIST) and demonstrate that it yields improved performance over other non-iterative methods and optimization-based methods.

\end{abstract}

\section{Introduction}
\label{sec:intro}
The two-dimensional discrete Fourier transform $\mathcal{F}(x)$ of an image
$x\in\mathbb{R}^{n\times n}$ can be represented by the magnitude $\omega$ and the phase $\varphi$, more precisely
\begin{align}
	\omega &= | \mathcal{F}(x)| \in \mathbb{R}^{n\times n},\label{eqn:measurement}\\
	\varphi&=\arg \mathcal{F}(x)\in [-\pi,
	\pi]^{n\times n},
\end{align}
where $\arg$ denotes the argument of a complex number (that is applied
element-wise).  Fourier phase retrieval is the problem of
reconstructing the original image only from its magnitude $\omega$.

While zero-padding is often assumed, it is a strong assumption on the support of $x$ which facilitates the phase retrieval problem.
Concretely, it assumes that we are reconstructing an $m\times m$ image
\begin{equation}
	\arraycolsep=5pt
	x_\text{padded} = \left[\begin{array}{cc} x &{0}_{n,m-n} \\ {0}_{m-n,n}  & {0}_{m-n,m-n} \end{array}\right]\in \mathbb{R}^{m\times m},
\end{equation}
where the ${0}_{a,b}$  denotes the $a\times b$ matrix with zeros.  The
oversampled magnitude can then be written as
\begin{equation}
	\omega_\text{oversampled} = |\mathcal{F}(x_\text{padded}) |\in \mathbb{R}^{m\times m}.
\end{equation}
For example, given $m=2n$, the magnitude is oversampled by a factor of four when
considering the two-dimensional case. There exist algorithms, e.g., the
Gerchberg-Saxton algorithm \cite{gerchberg1972practical} or Fienup's hybrid
input-output algorithm \cite{fienup1987phase}, that are able to reconstruct the
image from the magnitude that is oversampled by a factor of four. 
However, in practice the true images to be recovered
are not zero-padded and the magnitude is almost never oversampled. So the
assumption of zero-padding does not hold in general, as many applications measure
the non-oversampled magnitude  (i.e., $m = n$)  posing a great challenge for existing phase
retrievals methods.  
In this paper, we try to solve the more difficult problem, where
we reconstruct the image from the non-oversampled magnitude $\omega$.

\subsection{The Phase Contains the Relevant Information}

\begin{figure}
	\centering
	\def\arraystretch{7}
	\addtolength{\tabcolsep}{10pt}
	\begin{tabular}{cccc} 
		
		\multirow{2}{*}{\stackunder[8pt]{\includegraphics[width=2cm]{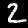}}{$x$}}&
		
		\stackunder[8pt]{\includegraphics[width=2cm]{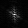}}{$\omega$}&
		\stackunder[8pt]{\includegraphics[width=2cm]{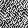}}{$\tilde\varphi$}&
		\stackunder[8pt]{\includegraphics[width=2cm]{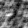}}{$x_{\tilde\varphi}$}\\
		&
		\stackunder[8pt]{\includegraphics[width=2cm]{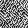}}{$\varphi$}&
		\stackunder[8pt]{\includegraphics[width=2cm]{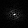}}{$\tilde\omega$}&
		\stackunder[8pt]{\includegraphics[width=2cm]{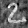}}{$x_{\tilde\omega}$}
	\end{tabular}
	\caption{Most information about the image is contained in the phase, which can be demonstrated by exchanging the phase with a random phase. For comparison we also exchange the magnitude with a random magnitude. Original image $x$, original magnitude $\omega$,  random phase $\tilde{\varphi}$, image obtained by combining the original magnitude and the random phase $x_{\tilde{\varphi}}$, original phase $\varphi$, random magnitude $\tilde \omega$, image obtained by combining the original phase and the random magnitude $x_{\tilde \omega}$.}
	\label{fig:phase}
\end{figure}

It is well known, that the phase contains most of the information of the image.
This can be observed by comparing an image with a random phase to an image with
a random magnitude.  To create these images we exchange (i) the phase of an
image by a random phase $\tilde \varphi$ which has entries that were uniformly sampled from
$[-\pi, \pi]$ while respecting the symmetries of the phase (to ensure a
real-valued image), and (ii) the magnitude with a random magnitude $\tilde \omega$ that has been sampled from
a truncated normal distribution with appropriate parameters.  To create an image
given the random phase $\tilde\varphi$ and the correct magnitude $\omega$, we apply the relationship
\begin{equation}
	x_{\tilde\varphi} = \mathcal{F}^{-1}\left(\omega \odot \exp(i\tilde\varphi)\right),
\end{equation}
where $\mathcal{F}^{-1}$ is the inverse Fourier transform, $i=\sqrt{-1}$ is the imaginary unit and $\odot$ is the
elementwise multiplication.
Analogously, we construct the image with the original phase $\varphi$ and a random magnitude $\tilde\omega$ as 
\begin{equation}
x_{\tilde\omega}  = \mathcal{F}^{-1}\left(\tilde\omega \odot \exp(i\varphi)\right).
\end{equation}
Fig.~\ref{fig:phase} shows that the image with the random phase is completely destroyed whereas the image with the random magnitude only exhibits some cloud-like artifacts.

\subsection{Non-Iterative Phase Retrieval}

To tackle the non-oversampled phase retrieval problem we formulate phase
retrieval as a learning problem.  Concretely, non-iterative phase retrieval
directly recovers the image from the magnitude only using a mapping that has
been learned to solve the problem in a particular problem domain. The mapping is
parameterized by a neural network $G$ that is trained to invert the measurement
process, i.e., 
\begin{equation}
	\hat{x} \approx G(\omega).
\end{equation}
Since the measurement process is known, training pairs can be generated
on-the-fly from sample images of a given dataset. The weights of $G$ can then be
learned using stochastic gradient descent by minimizing a loss function. The
benefit of non-iterative methods is the fast computation of the reconstruction
because only a single forward-pass through the neural network is used to
calculate the reconstruction.

\subsection{Contributions}
This paper addresses the challenge of improving the performance of non-iterative phase retrieval methods based on neural networks.  We show that a multi-scale approach based on cascading neural networks is able to improve previous non-iterative phase retrieval methods.

\subsection{Related Work}
Cascades of neural networks have been proposed previously by Schlemper et al. \cite{schlemper2017deep} but in the context of compressed sensing which is a related but different problem than phase retrieval. 
Phase retrieval has applications in many areas of research, e.g., in X-ray crystallography \cite{millane1990phase}, astronomical imaging \cite{fienup1987phase} or microscopy \cite{zheng2013wide}. 
We distinguish between three classes of methods for phase retrieval:

\begin{enumerate}
	\item Iterative methods without a learned component: Gerchberg and Saxton \cite{gerchberg1972practical} proposed a simple algorithm that is based on alternating reflections. 
	The idea behind this algorithm is to iteratively enforce the constraints in the Fourier space and the image space.
	Later Fienup modified the Gerchberg-Saxton algorithm in different ways which led to the input-output, the output-output and the hybrid-input-output (HIO) algorithm \cite{fienup1982phase}, where the HIO algorithm is most commonly used for phase retrieval. 
	Luke \cite{luke2004relaxed} analyzed the relaxed averaged alternating reflection (RAAR) algorithm.
	In general, these iterative methods without a learning component work well when the signal is oversampled.

	\item{Iterative methods with a learned component:}
	For non-oversampled phase retrieval Işıl et al. \cite{icsil2019deep} extend the HIO algorithm by a neural network that removes artifacts.
	Metzler et al. \cite{metzler2018prdeep} and Wu et al. \cite{wu2019online} use the regularization-by-denoising framework \cite{romano2017little} to solve oversampled phase retrieval problems.
	Another class of learned methods rely on the optimization of a latent variable of a learned generative model \cite{hand2018phase,uelwer2019phase} and produce high quality results.
	However, these methods require a training phase and an optimization phase during application and are therefore very costly.

	\item{Non-iterative methods with a learned component:}
	Non-iterative phase retrieval with a deep convolutional neural network that is trained end-to-end is proposed by Nishizaki et al. \cite{nishizaki2020analysis}.
	Recently, Tayal et al. \cite{manekar2020end} use symmetry breaking to solve the oversampled phase retrieval problem with neural networks.
	The benefit of non-iterative learned methods is the highly efficient reconstruction of images using only a single forward-pass through the model while also producing good results in the non-oversampled case.
\end{enumerate}

\section{Proposed Method}

\label{sec:method}
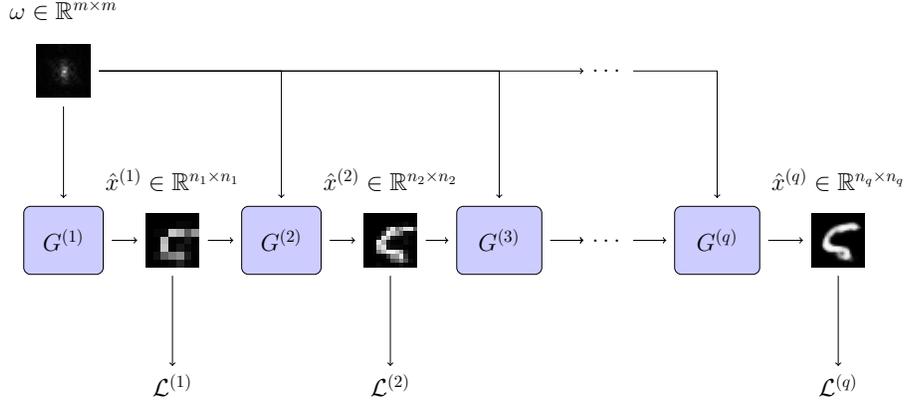
\begin{figure*}
	\centering
	\resizebox{\textwidth}{!}{
		\begin{tikzpicture}
			\node[inner sep=5pt] (magn) at (0,0) {\includegraphics[width=.09\textwidth]{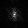}};
			
			\node[block] (G1) at (0,-3.5) {\Large $G^{(1)}$};
			\node[outer sep= 5, block] (G2) at (4.5,-3.5) {\Large $G^{(2)}$};
			\node[block, inner sep= 5] (G3) at (9,-3.5) {\Large $G^{(3)}$};
			\node[block, inner sep= 5] (Gq) at (13.5,-3.5) {\Large $G^{(q)}$};
			
			\node[inner sep=5] (dots) at (11.25, -3.5) {\Large $\dots$};
			\node[inner sep=5] (udots) at (11.25, 0) {\Large $\dots$};
			\node[inner sep=5] (L1) at (2.25,-6.5) {\Large $\mathcal{L}^{(1)}$};
			\node[inner sep=5] (L2) at (6.75,-6.5) {\Large $\mathcal{L}^{(2)}$};
			\node[inner sep=5] (Lq) at (16,-6.5) {\Large $\mathcal{L}^{(q)}$};
			\node[inner sep=5] (omega) at (0, 1.25) {\Large $\omega\in \mathbb{R}^{m\times m}$};
			\node[inner sep=5] (x) at (16, -2.25) {\Large $\hat{x}^{(q)}\in \mathbb{R}^{n_q\times n_q}$};
			\node[inner sep=5] (x1) at (2.25, -2.25) {\Large $\hat{x}^{(1)}\in \mathbb{R}^{n_1\times n_1}$};
			\node[inner sep=5] (x2) at (6.75, -2.25) {\Large $\hat{x}^{(2)}\in \mathbb{R}^{n_2\times n_2}$};
			
			\draw[->] (magn) edge (G1);
			\draw[->] (magn) -| (G2);
			\draw[->] (magn) -| (G3);
			\draw[->] (magn) -- (udots);
			\draw[->, ] (udots) -| (Gq);
			
			\node[inner sep=5pt] (r1) at (2.25,-3.5) {\includegraphics[width=.09\textwidth]{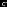}};
			\node[inner sep=5pt] (r2) at (6.75,-3.5) {\includegraphics[width=.09\textwidth]{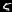}};
			\node[inner sep=5pt] (orig) at (16,-3.5) {\includegraphics[width=.09\textwidth]{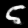}};
			
			\draw[->, ] (G1) - -(r1);
			\draw[->, ] (G2) -- (r2);
			\draw[->, ] (G3) -- (dots);
			\draw[->, ] (r1) -- (G2);
			\draw[->, ] (r2) -- (G3);
			\draw[->, ] (G3) -- (dots);
			\draw[->] (dots) -- (Gq);
			\draw[->, ] (Gq) -- (orig);
			
			\draw[->, ] (r1) -- (L1);
			\draw[->, ] (r2) -- (L2);
			\draw[->, ] (orig) -- (Lq);
	\end{tikzpicture}}
	\caption{An overview of the network architecture of the CPR approach. The magnitude image is fed to each of the networks. The sub-networks are updated stage-wise, i.e., we use $\mathcal{L}_1$ to update $G^{(1)}$, then the output of $G^{(1)}$ is passed as additional input to $G^{(2)}$ and so on. The first few networks focus on reconstructing a sub-sampled instance of the image, whereas the last sub-network predict the image at full-resolution.}
	\label{fig:network}
\end{figure*}

In this paper, we propose to use a cascaded neural network architecture for Fourier phase retrieval. Throughout the paper we refer to it as cascaded phase retrieval (CPR) network.
The CPR network consists of multiple sub-networks $G^{(1)}, \dots, G^{(q)}$ which are updated successively to reconstruct the different down-sampled instances of the original image, where  $G^{(2)}, \dots, G^{(q)}$ are fed with the intermediate reconstruction produced by the previous network.
In that way, each of these sub-networks can iteratively refine the reconstruction.
In addition to that, each of the sub-networks is provided with the measurement $\omega$ as an input.
The first few sub-networks are trained to reconstruct a down-sampled version of the image, where we denote the resolutions by $n_p\times n_p$ for $p = 1, \dots, q$. The last sub-networks predict the image at full-resolution $n_q\times n_q$.
The nearest-neighbor interpolation scheme is used for down-sampling the training images.
Fig.~\ref{fig:network} shows an overview of the CPR network architecture.

\subsection{Loss Functions}
A common choice for reconstruction tasks is the mean squared error (MSE) which can be defined for a batch $X = (x_1, \dots, x_b)$ of original images and a corresponding batch of reconstructions $\hat X = (\hat x_1, \dots, \hat x_b)$ as
\begin{equation}
	\label{eqn:mse}
	\mathcal{L}^{(p)}_\text{MSE}(X,\hat{X})= \dfrac{1}{b}\dfrac{1}{n_p^2}\sum_{k=1}^{b}\sum_{u=1}^{n_p} \sum_{v=1}^{n_p} \left(x_k[u, v]-\hat{x}_k[u, v]\right)^2.
\end{equation}
Although, it seems to work well in practice and provides good gradients for training, the reconstructions tend to be blurry.
This phenomenon has been discussed in \cite{pathak2016context}.
Hence, we also implement the mean absolute error (MAE), i.e., 
\begin{equation}
	\label{eqn:mae}
	\mathcal{L}^{(p)}_\text{MAE}(X,\hat{X})= \dfrac{1}{b}\dfrac{1}{n_p^2}\sum_{k=1}^b\sum_{u=1}^{n_p} \sum_{v=1}^{n_p}  \left|x_k[u,v]-\hat{x}_k[u,v]\right|
\end{equation} for measuring the reconstruction error.

\subsection{Training}
During training, each sub-network $G^{(p)}$  is trained using an individual loss $\mathcal{L}^{(p)}$. Each sub-network is updated one after another, where the loss $\mathcal{L}^{(p)}$ influences only $G^{(p)}$ and does not impact the parameters of the previous sub-networks.
Alternatively, the CPR network could be trained in an end-to-end fashion, however, since the intermediate reconstructions have different resolutions, we would need to carefully choose weights to balance the influence of each loss function $\mathcal{L}^{(1)}, \dots, \mathcal{L}^{(q)}$.
The training procedure is shown in more detail in Alg.~\ref{alg:train}.

\SetAlFnt{\normalsize}
\SetAlCapFnt{\normalsize}
\SetAlCapNameFnt{\normalsize}
\begin{algorithm}[ht]
	\label{alg:train}
	\setstretch{1.35}
	\SetKwInOut{Input}{Input}
	\Input {Dataset $X$, downsampling functions $g_2, \dots, g_q$, networks $G_1, \dots, G_q$, loss functions $\mathcal{L}^{(1)}, \dots, \mathcal{L}^{(q)}$}
	\For{$e \ =1,\ldots, N$}{
		\For{$\textup{batch } (x_1, \dots, x_b)\textup{ in }X$}{
		Calculate magnitudes $\Omega = (\omega_1, \dots, \omega_b)$ with $\omega_k = |\mathcal{F}(x_k)|,$ for $k=1,\dots, b$
		
		\For{ $p =1, \ldots, q$}{
			
			Calculate $\tilde{X}^{(p)} = (\tilde x_1, \tilde x_2, \dots, \tilde x_b)$, where $\tilde x_k = g_p(x_k)$ for $k=1,\dots, b$
			
			\uIf{$p==1$}{
				$\hat{X}^{(p)}  = G_p(\Omega)$
			}\uElse{

				$\hat{X}^{(p)} = G_p(\Omega, \hat X^{(p-1)} )$
			}
			Update network parameters using $\nabla \mathcal{L}^{(p)}\left(\hat{X}^{(p)} ,  \tilde X^{(p)}\right)$
		}
	}
	}
	\caption{Training algorithm for CPR network}
\end{algorithm}

\section{Experimental Evaluation}
\label{sec:experiments}
In this section, we empirically evaluate the performance of our model.
In order to do this, we report the results of the fully-convolutional residual network (ResNet) employed by Nishizaki et al. \cite{nishizaki2020analysis}, the  multi-layer-perceptron (MLP) used in \cite{uelwer2019phase} and the PRCGAN \cite{uelwer2019phase}.
In addition to these learned networks we include the results of the well-established HIO algorithm \cite{fienup1982phase} and the RAAR algorithm \cite{luke2004relaxed} as a baseline.

\subsection{Datasets}
For the experimental evaluation we use the MNIST \cite{mnist}, the EMNIST \cite{cohen2017emnist}, the Fashion-MNIST \cite{fashion} and the KMNIST \cite{clanuwat2018deep} datasets.
All datasets consist of $28\times 28$ grayscale images, i.e., $n=28$. 
MNIST contains images of digits, EMNIST contains images of letters and digits, Fashion-MNIST contains images of clothing and KMNIST contains images of cursive Japanese characters.
Although these datasets are considered to be toy datasets when it comes to classification tasks, they provide quite challenging data for two-dimensional Fourier phase retrieval.
For the EMNIST dataset we use the balanced version of the dataset.

\subsection{Experimental Setup}

We compare our CPR approach with the MLP and the ResNet that are trained to minimize $\mathcal{L}_\text{MSE}$ for the MNIST, the EMNIST and the KMNIST dataset.
The  $\mathcal{L}_\text{MAE}$ is used for the Fashion-MNIST dataset. 
Furthermore, we report the results of an MLP trained with an adversarial loss in combination with $\mathcal{L}_\text{MAE}$ (PRCGAN) as proposed in \cite{uelwer2019phase}.
For our proposed CPR network we consider a cascade of five MLPs with three hidden layers where we increased the scales of the (intermediate) reconstructions according to Tab.~\ref{tab:number-of-hiddens}. The number of hidden units for each sub-network is also shown in Tab.~\ref{tab:number-of-hiddens}. 
Furthermore, we compare the results with a CPR network that produces intermediate reconstructions at full scale. We refer to this variant as CPR-FS.
All sub-networks are trained using dropout \cite{srivastava2014dropout}, batch-normalization \cite{ioffe2015batch} and ReLU activation functions. For the last layer we use a Sigmoid function to ensure that the predicted pixels are in $[0,1]$. To optimize the weights we used Adam \cite{kingma2014adam} with learning rate $10^{-4}$.
We train all versions of the CPR network for $100$ epochs with the $\mathcal{L}_\text{MSE}$, except for the Fashion-MNIST dataset where we use $\mathcal{L}_\text{MAE}$ for the final layer.
These choices gave the best results on the validation dataset.

We ran the HIO algorithm and the RAAR algorithm for $1000$ steps each and allowed three random restarts, where we selected the reconstruction $\hat{x}$ with the lowest magnitude error $|||\mathcal{F}(\hat{x})|-\omega||_\text{Fro}$.
For HIO we set $\beta=0.8$ and for  RAAR we set $\beta=0.87$.

\begin{table}
	\caption{Scales used for the (intermediate) reconstructions and number of hidden units used for each network of the cascade.}
	\label{tab:number-of-hiddens} 
	\centering
	\addtolength{\tabcolsep}{3.5pt}
	\begin{tabular}{llccccc} 
		\toprule
		&   &$G^{(1)}$ &$G^{(2)}$ & $G^{(3)}$ & $G^{(4)}$& $G^{(5)}$\\  
		\midrule
		\multirow{2}{*}{Scale} &CPR  & $7\times 7$ & $12\times 12$ & $17\times 17$ &  $22\times 22$ &  $28\times 28$\\
		& CPR-FS  & $28\times 28$ & $28\times 28$ & $28\times 28$ &  $28\times 28$ &  $28\times 28$\\
		\midrule
		\multirow{2}{*}{Hidden layer size} & CPR  & $1136$ & $1336$ & $1536$ & $1736$ &  $ 1936$ \\
		& CPR-FS &  $1936$ & $1936$ & $1936$ & $1936$ &  $ 1936$ \\
		\bottomrule
	\end{tabular}
\end{table}

\subsection{Metrics}
For a quantitative evaluation we compare the MSE and the MAE as defined in Eq.~\ref{eqn:mse} and Eq.~\ref{eqn:mae}.
Moreover, we report the structural similarity index (SSIM) that was introduced by Wang et al. \cite{wang2004image}. 
The SSIM measures perceived quality of an reconstruction on various windows of an image and takes values between $0$ (worst quality) and $1$ (perfect reconstruction).

Because translating signals by a constant shift or rotating them by $180\degree$ does not change their Fourier magnitude, we considered these reconstructions equally correct.
Thus, we register the predictions (and their rotated variants) using cross-correlation as described by Brown \cite{brown1992survey} before calculating the evaluation metrics.

\subsection{Results}

\begin{figure}
	\centering
	\addtolength{\tabcolsep}{5pt}
	\begin{tabular}{rcc} 
		& MNIST & Fashion-MNIST\\
		HIO \cite{fienup1982phase}& \includegraphics[height=0.72cm, align=c]{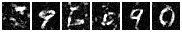} & \includegraphics[height=0.72cm, align=c]{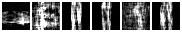}\vspace{0.1cm}\\
		RAAR \cite{luke2004relaxed} & \includegraphics[height=0.72cm, align=c]{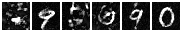}& \includegraphics[height=0.72cm, align=c]{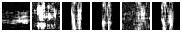}\vspace{0.1cm}\\
		ResNet \cite{nishizaki2020analysis} & \includegraphics[height=0.72cm, align=c]{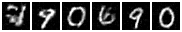}& \includegraphics[height=0.72cm, align=c]{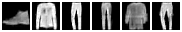}\vspace{0.1cm}\\
		MLP \cite{uelwer2019phase}& \includegraphics[height=0.72cm, align=c]{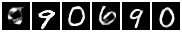}& \includegraphics[height=0.72cm, align=c]{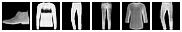}\vspace{0.1cm}\\
		PRCGAN \cite{uelwer2019phase}  & \includegraphics[height=0.72cm, align=c]{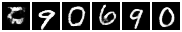}& \includegraphics[height=0.72cm, align=c]{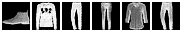}\vspace{0.1cm}\\
		CPR (ours)&\includegraphics[height=0.72cm, align=c]{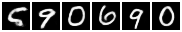}&\includegraphics[height=0.72cm, align=c]{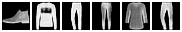}\vspace{0.1cm}\\
		CPR-FS (ours)&\includegraphics[height=0.72cm, align=c]{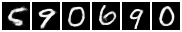}&\includegraphics[height=0.72cm, align=c]{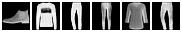}\vspace{0.1cm}\\
		Original & \includegraphics[height=0.72cm, align=c]{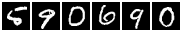}& \includegraphics[height=0.72cm, align=c]{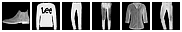}\\
	\end{tabular}
	\caption{Reconstructions from the Fourier magnitudes of samples from the MNIST and the Fashion-MNIST test dataset.}
	\label{fig:MNIST}
\end{figure}

Fig.~\ref{fig:MNIST} compares six reconstructions by the different methods on the MNIST and the Fashion-MNIST test dataset.
We observe that the HIO algorithm and the RAAR algorithm fail to recover the image in most of the cases.
From all learned methods, the Resnet produced the worst reconstructions. The estimated images are very blurry and in some cases the reconstruction exhibit deformations (e.g., the last two images from the Fashion-MNIST dataset that are shown in Fig.~\ref{fig:MNIST}).
The PRCGAN produces reconstructions that are sharp and overall the visual quality is similar to the reconstructions of the  MLP.
Most of the learned methods struggle to recover the first image of the MNIST dataset (depicting the "$5$") . 
We suppose that this sample is very different from the  samples that were used to train the networks.
Only, the CPR and the CPR-FS network are capable of recovering this image.

\begin{figure}
	\centering
	\includegraphics[width=0.8\linewidth]{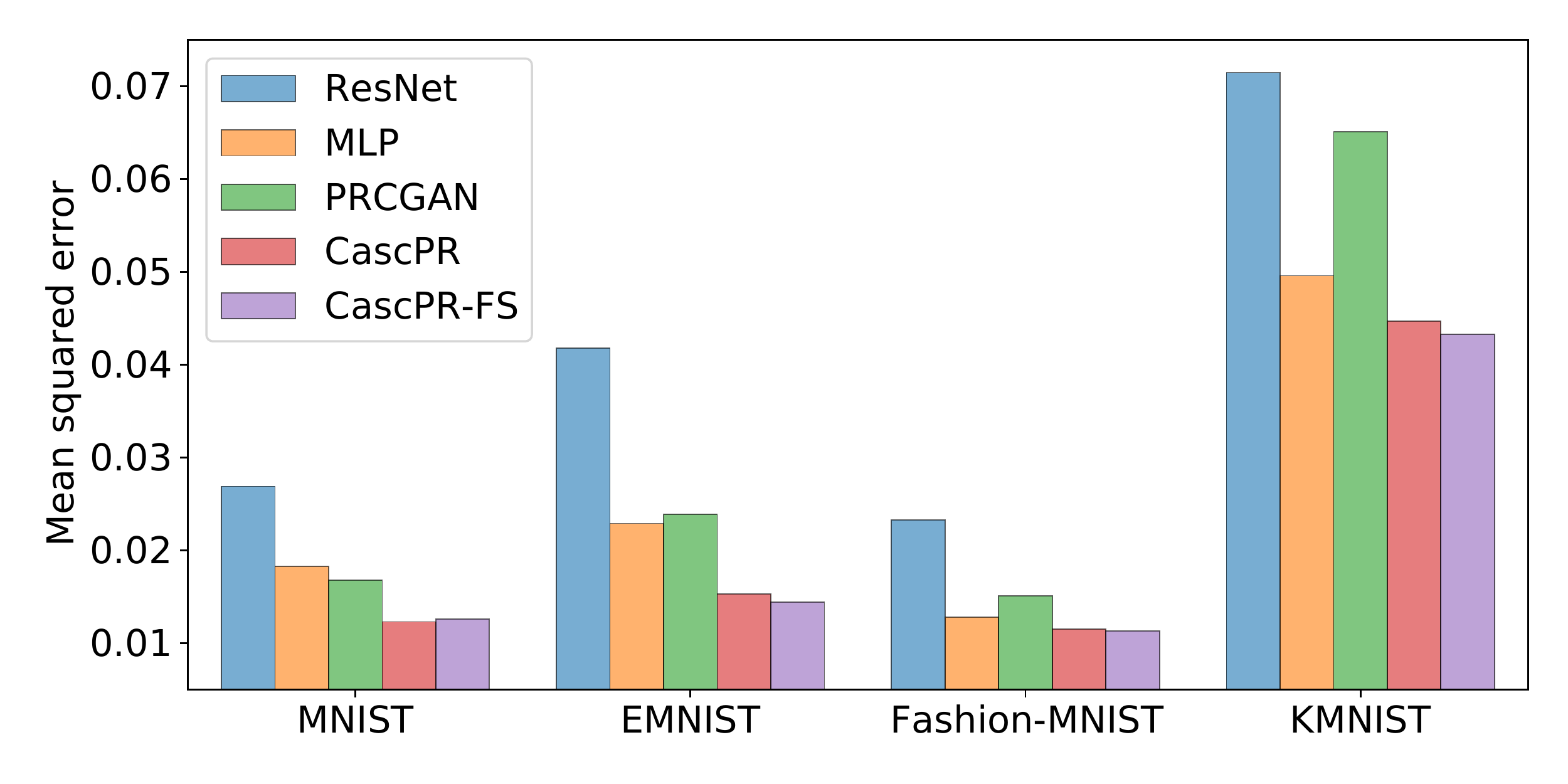}
	\caption{ Comparison of the MSE for the results of the learned methods.}
	\label{fig:mse}
\end{figure}

Tab.~\ref{tab:results} shows the MSE, the MAE and the SSIM of the reconstructions and Fig.~\ref{fig:mse} visualizes the MSE for the five different learned methods.
Overall, the learned methods outperform RAAR and HIO by a large margin.
For MNIST, EMNIST and KMNIST we see that the CPR network greatly improves the reconstruction quality compared to the other learned methods.
We hypothesize that our proposed CPR network yields better results when the signals of interest have a small support (e.g., MNIST, EMNIST, KMNIST).
However, for signals with a large support (e.g., Fashion MNIST) we only observe a small improvement compared to the other learned methods.

\begin{table*}
	\caption{Quantitative comparison of the reconstructions produced by the different methods. We report MSE, MAE and SSIM between the reconstructions and the original images of the test dataset. MSE, MAE: lower is better. SSIM: larger is better. The best result is printed \textbf{bold}.}
	\label{tab:results} 
	\centering
	\addtolength{\tabcolsep}{5pt}
	\begin{tabular}{l ccc ccc} 
		\toprule
		&\multicolumn{3}{c}{MNIST}&\multicolumn{3}{c}{EMNIST}\\ 
		& MSE  & MAE & SSIM  & MSE & MAE & SSIM \\ 
		\midrule
		HIO \cite{fienup1982phase} & 0.0441 & 0.1016 & 0.5708 & 0.0653 &  0.1379 & 0.5241\\ 
		RAAR \cite{luke2004relaxed} & 0.0489 & 0.1150 & 0.5232 & 0.0686 & 0.1456 & 0.4973 \\
		ResNet \cite{nishizaki2020analysis} & 0.0269 & 0.0794 & 0.6937 & 0.0418 & 0.1170 & 0.5741 \\
		MLP \cite{uelwer2019phase} & 0.0183 & 0.0411 & 0.8345 & 0.0229 & 0.0657 & 0.7849 \\ 
		PRCGAN \cite{uelwer2019phase} & {0.0168 }& 0.0399 & 0.8449 & 0.0239 & 0.0601 & 0.8082 \\ 
		CPR (ours) & \textbf{0.0123} & \textbf{0.0370} & \textbf{0.8756 }& 0.0153 & 0.0525 & 0.8590 \\
		CPR-FS (ours) & 0.0126 & 0.0373&0.8729 & \textbf{0.0144}& \textbf{0.0501}& \textbf{0.8700} \\ 
		\midrule& \multicolumn{3}{c}{Fashion-MNIST}&\multicolumn{3}{c}{KMNIST}\\ 
		& MSE & MAE & SSIM & MSE & MAE & SSIM \\ 
		\midrule
		HIO \cite{fienup1982phase} & 0.0646 & 0.1604 & 0.4404 &  0.0835 & 0.1533 & 0.3414\\ 
		RAAR \cite{luke2004relaxed} & 0.0669 & 0.1673 & 0.4314 & 0.0856  & 0.1559 & 0.3208\\
		ResNet \cite{nishizaki2020analysis} & 0.0233 & 0.0820 & 0.6634 & 0.0715  &  0.1711 & 0.3783\\
		MLP \cite{uelwer2019phase}  & 0.0128 & 0.0526 & 0.7940 & 0.0496 & 0.1168 & 0.5991\\ 
		PRCGAN \cite{uelwer2019phase} & 0.0151 & 0.0572 & 0.7749 & 0.0651 & 0.1166 & 0.5711\\ 
		CPR (ours) &0.0115 & 0.0503 & 0.8077& 0.0447 & 0.1068  & 0.6488 \\
		CPR-FS (ours) &\textbf{0.0113 }& \textbf{0.0497}& \textbf{0.8092} & \textbf{0.0433} & \textbf{0.1034}  & \textbf{0.6626} \\ 
		\bottomrule
	\end{tabular}
\end{table*}

\subsection{Intermediate Prediction at Full-Scale}

We briefly study the effect of predicting down-sampled versions of the image. 
Therefore, we evaluate the CPR-FS network which produces full-scale intermediate reconstructions.
Tab.~\ref{tab:results} also shows that the CPR-FS network performs similarly in terms of the overall reconstruction quality. 
For the EMNIST, the Fashion-MNIST and the KMNIST dataset the full-scale variant is slightly better.
However, due to the larger input, the sub-networks need to have more parameters and thus training is more expensive.

\subsection{Ablation Study}
\begin{figure}
 	\centering
 	\includegraphics[width=0.8\linewidth]{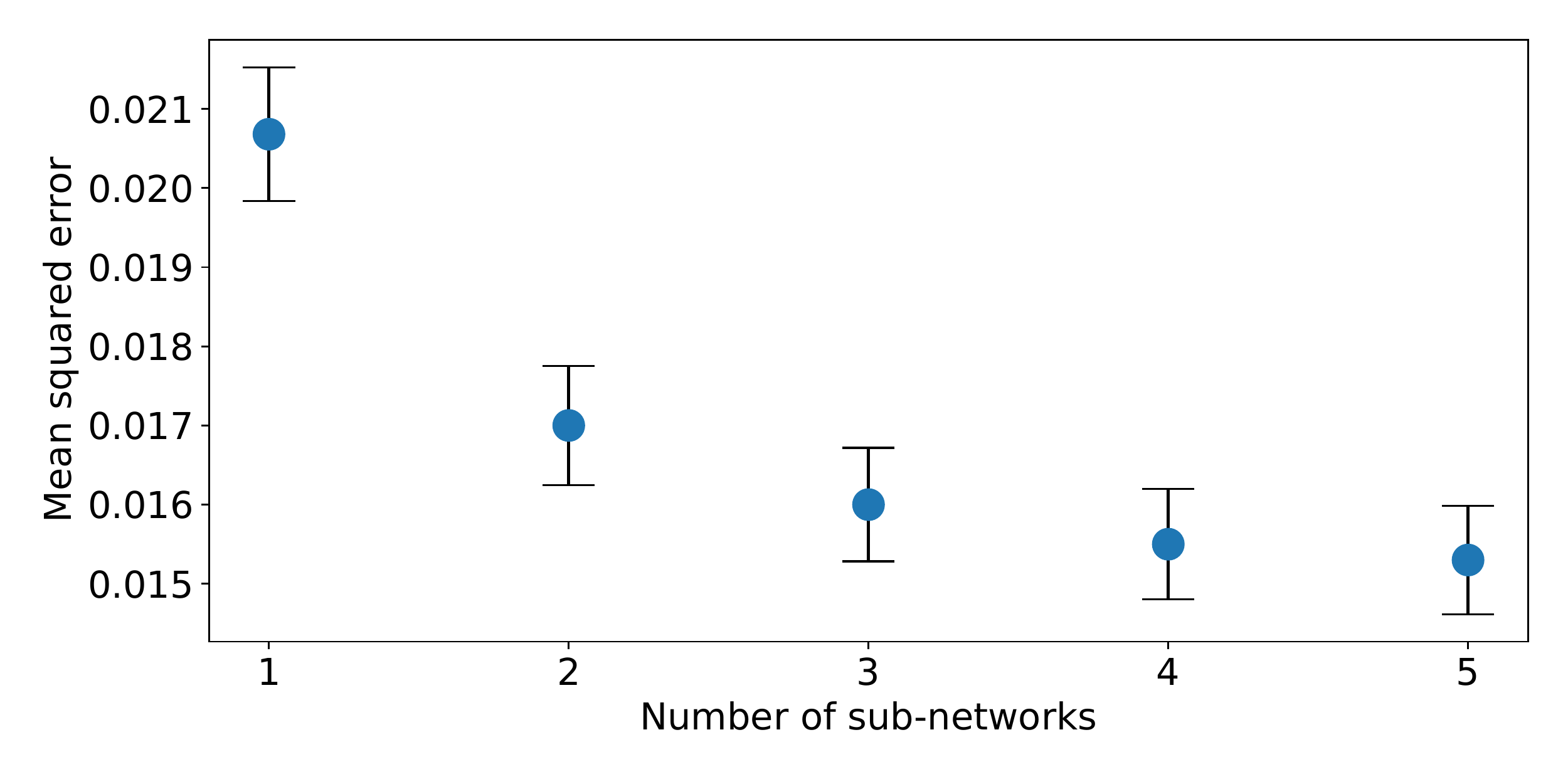}
 	\caption{Test MSE on the EMNIST test dataset for different number of sub-networks. Error bars indicate the 95\% confidence interval.  }
 	\label{fig:error}
 \end{figure}

In this section, we demonstrate that increasing the number of sub-networks has a beneficial effect on the overall reconstruction quality.
To do so, we train five network cascades exemplarily on the EMNIST dataset where we increase the number of sub-networks from one to five.
We report the MSE on the test dataset after $50$ epochs.
Fig.~\ref{fig:error} shows that the MSE for the EMNIST dataset decreases with an increasing number of sub-networks used for the CPR-FS approach. 
Furthermore the gain in terms of MSE saturates after $q=5$, such that additional sub-networks do not bring any further improvements.
 We expect the same relative behavior on the other datasets when increasing $q$.
 
\section{Conclusion and Future Work}
In this paper, we use a cascade of neural networks for non-oversampled Fourier phase retrieval.
Our approach successively reconstructs images from their Fourier magnitudes and outperforms other existing non-iterative networks noticeably in terms of the reconstruction quality.
However, non-iterative methods do not yet reach the reconstruction quality of iterative methods with a learning component which require high computational cost at test time.

Future work could also evaluate different strategies for training the neural network cascade. 
For example, greedy sub-network-wise training could be implemented and compared with our training procedure.
Moreover, the CPR network architecture can easily be adapted to solve inverse problems other than Fourier phase retrieval.

\bibliographystyle{splncs04}
\bibliography{bibliography}

\end{document}